\documentclass[a4paper,7pt,twoside]{CoAst}
\usepackage{epsf,graphicx,natbib,fancyhdr}

\renewcommand{\chaptermark}[1]{\markboth{#1}{}}

\newcommand{\aap}{A\&A}  

\newcommand{\kopf}{\small\itshape Comm. in Asteroseismology \\ Vol. number, publication date (will be inserted in the production process)}

\newcommand{\Authors}[1]{\begin{center}\normalsize\bf\sf #1 \end{center}}

\renewcommand{\author}[1]{\begin{center}\normalsize\bf\sf #1 \end{center}}
\newcommand{\Address}[1]{\begin{center}\small\sf #1 \end{center}}

\renewenvironment{abstract}{\section*{Abstract}\normalsize\sf}{}

\newcommand{\chapterCoAst}[3]{\chapter[\sf\normalsize #1\\ \footnotesize \hspace*{5mm}by #2 \sf\normalsize][]{#1\\}\rhead[\fancyplain{}{\sf\footnotesize \center{#3}}]{\fancyplain{}{\sffamily\thepage}}\lhead[\fancyplain{\kopf}{\sffamily\thepage}]{\fancyplain{\kopf}{\sf\footnotesize \center{#2}}}}

\renewcommand{\chaptername}{}

\newcommand{\acknowledgements}[1]{\vspace*{5mm}\noindent\begin{bf}Acknowledgments. \end{bf} #1}

\usepackage[margin=3cm,includehead,includefoot,heightrounded,a4paper]{geometry}

\bibpunct[; ]{(}{)}{;}{a}{}{}

\newcommand{\Dnu}{\mbox{$\Delta \nu$}}
\newcommand{\dnu}[1]{\mbox{$\delta \nu_{#1}$}}
\newcommand{\acena}{\mbox{$\alpha$~Cen~A}}
\newcommand{\acenb}{\mbox{$\alpha$~Cen~B}}

\newcommand{\eboo}{\mbox{$\eta$~Boo}}
\newcommand{\tcet}{\mbox{$\tau$~Cet}}
\newcommand{\muHz}{\mbox{$\mu$Hz}}

\newcommand{\new}[1]{{\bf #1}}
\renewcommand{\new}[1]{{#1}}
\newcommand{\half}{{\textstyle\frac{1}{2}}}

\newcommand{\gcm}{\mbox{g\,cm$^{-3}$}}

\begin{document}
\sf
  
\chapterCoAst{Scaled oscillation frequencies and \'echelle diagrams as a
    tool for comparative asteroseismology} 
{ T. R. Bedding and H. Kjeldsen }
{Scaled oscillation frequencies and \'echelle diagrams 
    for comparative asteroseismology} 

\Authors{Timothy R. Bedding$^1$ and Hans Kjeldsen$^{2}$} 
\Address{$^1$ Sydney Institute for Astronomy (SIfA), School of Physics,
  University of Sydney, NSW 2006, Australia\\ 
$^2$ Danish AsteroSeismology Centre (DASC), Department of Physics and
  Astronomy, Aarhus University, DK-8000 Aarhus C, Denmark}

\noindent
\begin{abstract} 
We describe a method for comparing the frequency spectra of oscillating
stars.  We focus on solar-like oscillations, in which mode frequencies
generally follow a regular pattern.  On the basis that oscillation
frequencies of similar stars scale homologously, we show how to display two
stars on a single \'echelle diagram.  The result can be used to infer the
ratio of their mean densities very precisely, without reference to
theoretical models.  In addition, data from the star with the better
signal-to-noise ratio can be used to confirm weaker modes and reject
sidelobes in data from the second star.  Finally, we show that scaled
\'echelle diagrams provide a solution to the problem of ridge identification
in F-type stars, such as those observed by the CoRoT space mission.
\end{abstract}


\section{Introduction}

This paper discusses how to compare frequency spectra of oscillating stars.
We focus on solar-like p-mode oscillations, in which mode frequencies
generally follow a regular pattern.  This makes it useful to characterize
them by a handful of frequency separations: the so-called large separation
\Dnu\ between consecutive overtones of a given angular degree~$l$, and the
small separations between adjacent modes of different degree.  These
frequency separations have the advantage of being closely related to
physical properties of the stellar interior (see Section~\ref{sec:asym}).
Measuring them and their variations with frequency and comparing with
theoretical models is a major focus of asteroseismology (see, for example,
reviews by \citealt{B+G94} and \citealt{ChD2004}).

There is somewhat less focus on absolute frequencies, partly because
stellar models \new{do not} properly model the near-surface layers
\citep{ChDDL88,DPV88,RChDN99,LRD2002}.  This makes it difficult to compare
individual observed frequencies with models, although \citet{KBChD2008}
have proposed an empirical correction that appears promising, at least for
stars reasonably close in effective temperature to the Sun.

Here, we wish to compare observations of one star with observations of
another, and so difficulties with models are not relevant.  We are
motivated by the expectation from homology that if two stars are
sufficiently similar, their oscillation frequencies will be in the same
ratio as the square roots of their mean densities:
\begin{equation}
  \frac{\nu_1}{\nu_2} = \sqrt{\frac{\bar{\rho}_1}{\bar{\rho}_2}}. \label{eq.ratio}
\end{equation}
Here, we are comparing modes in the two stars with the same radial order
($n$) and angular degree~($l$).  Even if the two stars are not similar, we
might still expect Equation~\ref{eq.ratio} to provide a useful
approximation.  \new{Of course, it also follows that the 
large separation scales in the same way:
\begin{equation}
  \frac{\Dnu_1}{\Dnu_2} = \sqrt{\frac{\bar{\rho}_1}{\bar{\rho}_2}}. \label{eq.ratio.Dnu}
\end{equation}}
We now present some examples and applications, using
\'echelle diagrams to visualize the comparisons between stars.

\section{\'Echelle diagrams and the asymptotic relation}
\label{sec:asym}

The \'echelle diagram, first introduced by \citet{GFP83} for global
helioseismology, is nowadays used extensively in asteroseismology as a
valuable way of displaying oscillation frequencies.  It involves dividing
the spectrum into segments of length \Dnu\ and stacking them one above the
other so that modes with a given degree align vertically in ridges.  Any
departures from regularity, such as variations in the large separation with
frequency, are clearly visible as curvature in the \'echelle diagram, and
variations in the small separations appear as a convergence or divergence
of the corresponding ridges.

We conventionally define three observable small frequency separations:
$\dnu{02}$ is the spacing between $l=0$ and $l=2$; $\dnu{13}$ is the
spacing between $l=1$ and $l=3$; and $\dnu{01}$ is the amount by which
$l=1$ is offset from the midpoint of the $l=0$ modes on either side.  In
practice, the large and small separations are observed to vary with
frequency.

The regularity in solar-like oscillation spectra allows us to write the
mode frequencies in terms of the large and small separations, as follows:
\begin{equation}
  \nu_{n,l} = \Dnu (n + \half l + \epsilon) - d_l,
        \label{eq.asymptotic}
\end{equation}
where $\epsilon$ is a dimensionless offset.  The small separation $d_l$ is
zero for $l=0$ (radial modes), and equals $\dnu{01}$ for $l=1$, $\dnu{02}$
for $l=2$ and $(\dnu{01}+\dnu{13})$ for $l=3$.  \citet{BHS2010} have
suggested that this last separation should be designated \dnu{03}.

Equation~\ref{eq.asymptotic} describes the oscillation frequencies from an
observational perspective.  A theoretical asymptotic expression
\citep{Tas80,Gou86,Gou2003} relates \Dnu, $d_l$ and $\epsilon$ to integrals
of the sound speed.  In particular, \Dnu\ measures quite accurately the
sound travel time across the diameter of the star, while the small
separations \new{are sensitive to the structure of the core} and
$\epsilon$ is sensitive to the surface layers.

When making an \'echelle diagram, it is common to plot $\nu$ versus ($\nu
\bmod \Dnu$), in which case each order slopes upwards slightly.  However,
for grayscale images it can be preferable to keep the orders horizontal.
We have done that in this paper, and so the value given on the vertical
axis is actually the frequency at the middle of the order.

\section{Scaled \'echelle diagrams and their applications}

The ridges in an \'echelle diagram will only appear vertical if we use the
correct value of \Dnu.  For this reason, it does not generally make sense
to plot two stars on the same \'echelle diagram.  However, if the frequencies
of the second star have been scaled by multiplying them all by the ratio of
the large separations, we are led by Equation~\ref{eq.ratio} to expect that
its ridges can be made to coincide with those of the first.  The scaling
factor can be fine-tuned to optimize the alignment in two different ways:
\begin{enumerate}

\item by matching the slopes for the two stars (making both vertical),
  which means matching $\Dnu$, or

\item by overlaying the ridges as closely as possible, although this may
  mean they have different slopes, which means forcing them to have the
  same value of~$\epsilon$.
\end{enumerate}

\begin{figure}[t]
\centering
\includegraphics*[width=90mm]{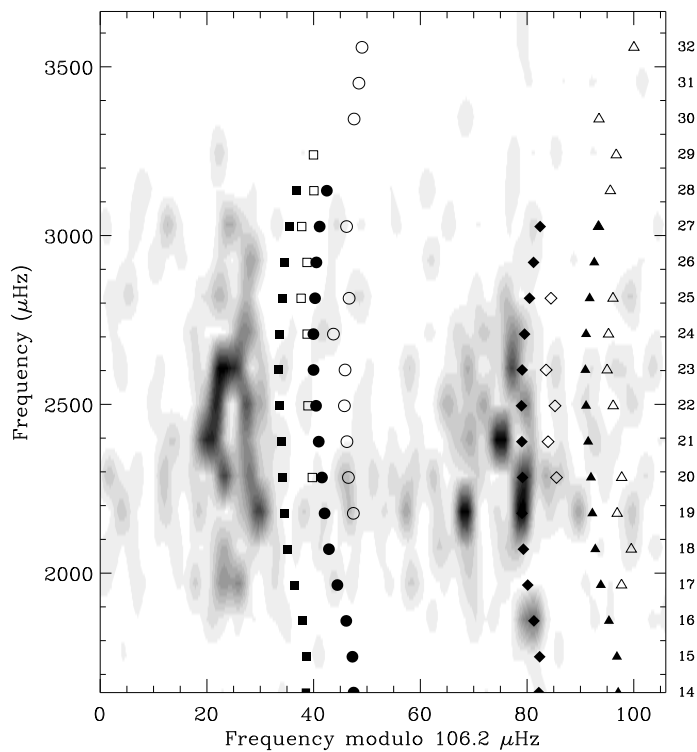}
\caption{Scaled \'echelle diagram comparing three main-sequence stars.  The
greyscale is the power spectrum of \acena, filled symbols are oscillation
frequencies for the Sun after multiplying by 0.7816, and open symbols are
frequencies for \acenb\ after multiplying by 0.6555 (see text for
references).  The scaling factors were fine-tuned to make the ridges for
all three stars parallel (Method~1).  Symbol shapes indicate mode degree:
$l=0$ (circles), $l=1$ (triangles), $l=2$ (squares) and $l=3$ (diamonds).}
\label{fig:acena-sun-acenb}
\end{figure}

Method~1 is shown in Figure~\ref{fig:acena-sun-acenb} for three
main-sequence stars: \acena\ ($\Dnu\approx 106\,\muHz$), the Sun
($\Dnu\approx 135\,\muHz$) and \acenb\ ($\Dnu\approx 162\,\muHz$).  The
greyscale is the power spectrum of \acena\ as observed in a two-site
campaign with UVES and UCLES \citep{BKB2004}, but with weights optimized to
minimize the sidelobes \citep{AKB2010}.  The filled symbols are oscillation
frequencies for the Sun \citep[][Table~2]{BCD2009} after multiplying by
0.7816, and open symbols are frequencies for \acenb\ \citep{KBB2005} after
multiplying by 0.6555.  The scaling factors were tuned \new{(using simple
`trial and error')} to make the ridges for all three stars parallel.  Note
that scatter of observations about the smooth ridges for \acena\ and~B is
due to the relatively short duration of the observations (only a few times
longer than the mode lifetimes).

There is a systematic progression in stellar parameters (mass, effective
temperature and luminosity) as we go from \acena\ through the Sun to
\acenb.  We see in Fig.~\ref{fig:acena-sun-acenb} a corresponding
progression in the positions of the ridges, which corresponds to a change
in~$\epsilon$ (see Equation~\ref{eq.asymptotic}).  Apart from this, there
is a close similarity between the three stars, although there are subtle
differences in the curvatures of the ridges and in the small separations
between them.

\begin{figure}[t]
\centering
\includegraphics*[width=90mm]{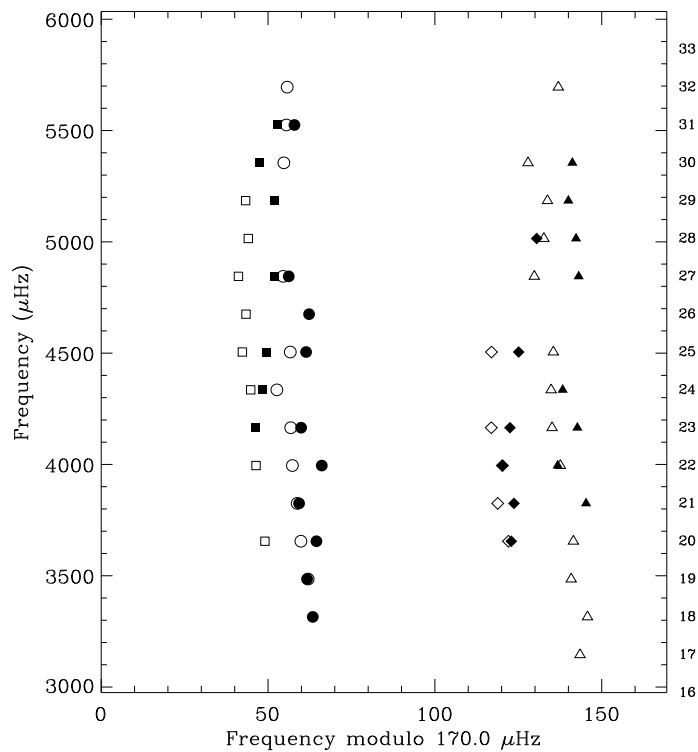}
\caption{Scaled \'echelle diagram comparing two simliar stars low-mass
stars.  Filled symbols: frequencies (unscaled) for \tcet.  Open symbols:
frequencies for \acenb{} multiplied by 1.0478. The scaling factor was tuned
to make the ridges coincide (Method~1/2).  Symbol shapes indicate mode
degree: $l=0$ (circles), $l=1$ (triangles), $l=2$ (squares) and $l=3$
(diamonds).  }
\label{fig:tcet-acenb}
\end{figure}

If two stars have very similar parameters, the ridges will almost coincide
(Methods~1 and 2 become the same).  This is the case for the Sun and the
solar twin 18~Sco, recently observed with HARPS and SOPHIE (M. Bazot et
al., in prep.), for which the scaling factor gives an extremely precise
measurement of the mean density.  In this case, we can also use one star as
a guide when identifying modes in the other (and eliminating aliases).
Another example is shown in Figure~\ref{fig:tcet-acenb}, for a pair of
low-mass stars: the filled symbols show observed frequencies for \tcet\
($\Dnu\approx 170\,\muHz$; \citealt{TKB2009}) and the open filled symbols
show those of \acenb{} \citep{KBB2005} after multiplying by 1.0478.

\subsection{Measuring mean densities}

{Assuming that a given pair of stars are homologous, (i.e., that
Equation~\ref{eq.ratio} holds)}, the value of the scaling factor gives a
direct measurement of their relative densities, without the need to refer
to theoretical models.  The scaling factors used in
Fig.~\ref{fig:acena-sun-acenb} are precise to about 0.05\%, in the sense
that changing them by this amount produces a noticable deviation from
parallelism.  If we accept the validity of Equation~\ref{eq.ratio} then our
results allow us to measure mean densities for \acena\ and~B relative to
solar with a precision of 0.1\%.  We obtain $(0.8601\pm 0.0003)\,\gcm$ for
\acena\ and $(2.0018\pm0.0008)\,\gcm$ for \acenb.  These values agree with
those found by comparing the observed frequencie of radial modes with
models that have been corrected for the near-surface offset
\citep[][Table~2]{KBChD2008}, but are more precise and do not make any use
of model calculations.  In practice, however, Equation~\ref{eq.ratio} may
not be accurate to this level.  {In particular, defining a mean density
requires that we specify the position of the surface, which is ambiguous,
as discussed by \citet{B+U88} in the context of helioseismology.  } In any
case, we have certainly derived mean densities for these stars that are
precise enough for any practical application.

The two stars shown in Figure~\ref{fig:tcet-acenb}, \tcet\ and \acenb, are
even closer.  Once again, we found that changing the scaling factor by
about 0.05\% produced a noticable departure from parallelism.  Using the
mean density found above for \acenb, the implied mean density for \tcet\ is
$2.198 \pm 0.004\,\gcm$, which agrees with the value of $2.21 \pm
0.01\,\gcm$ found by \citet{TKB2009} but is more precise.  {Again, we
note that being able to measure the homology scaling factor to high
precision does not necessarily provide a density measurement with similar
accuracy.}

\subsection{Ridge identification in F stars}
\label{sec:f-stars}

An important application of scaled \'echelle diagrams is to the problem of
ridge identification in F stars.  This problem has arisen in the context of
several F-type main-sequence stars observed using the CoRoT spacecraft.
The first and best-studied example is HD~49933 ($\Dnu \approx 85\,\muHz$),
whose \'echelle diagram from 60 days of CoRoT observations showed two broad
and very similar ridges \citep{AMA2008}.  It was clear that one ridge was
due to $l=0$ and $l=2$ modes and the other to $l=1$, but the combination of
significant rotational splitting and large linewidths made it difficult to
decide which was which.  \citet{AMA2008} made a global fit to the line
profiles, which led them to favor the possibility they labelled
`Scenario~A'.  {Further analysis of the same data has been carried out
by several groups \citep{BAB2009,GKW2009,Rox2009} and none favored a
definite identification, while comparison with theoretical models
\citep{KGG2010} gave a better match to Scenario~B.}  Subsequently, the
analysis of an additional 180 days of CoRoT observations using revised
methods led \citet{BBC2009} to reverse the original identification in favor
Scenario~B.

Two other F stars observed by CoRoT have presented the same problem, namely
HD~181906 ($\Dnu\approx 87.5\,\muHz$; \citealt{GRS2009}) and HD~181420
($\Dnu\approx 75\,\muHz$; \citealt{BDB2009}).  In neither case were the
authors able to decide the correct scenario.

\begin{figure}[p]
\centering \includegraphics*[width=72mm]{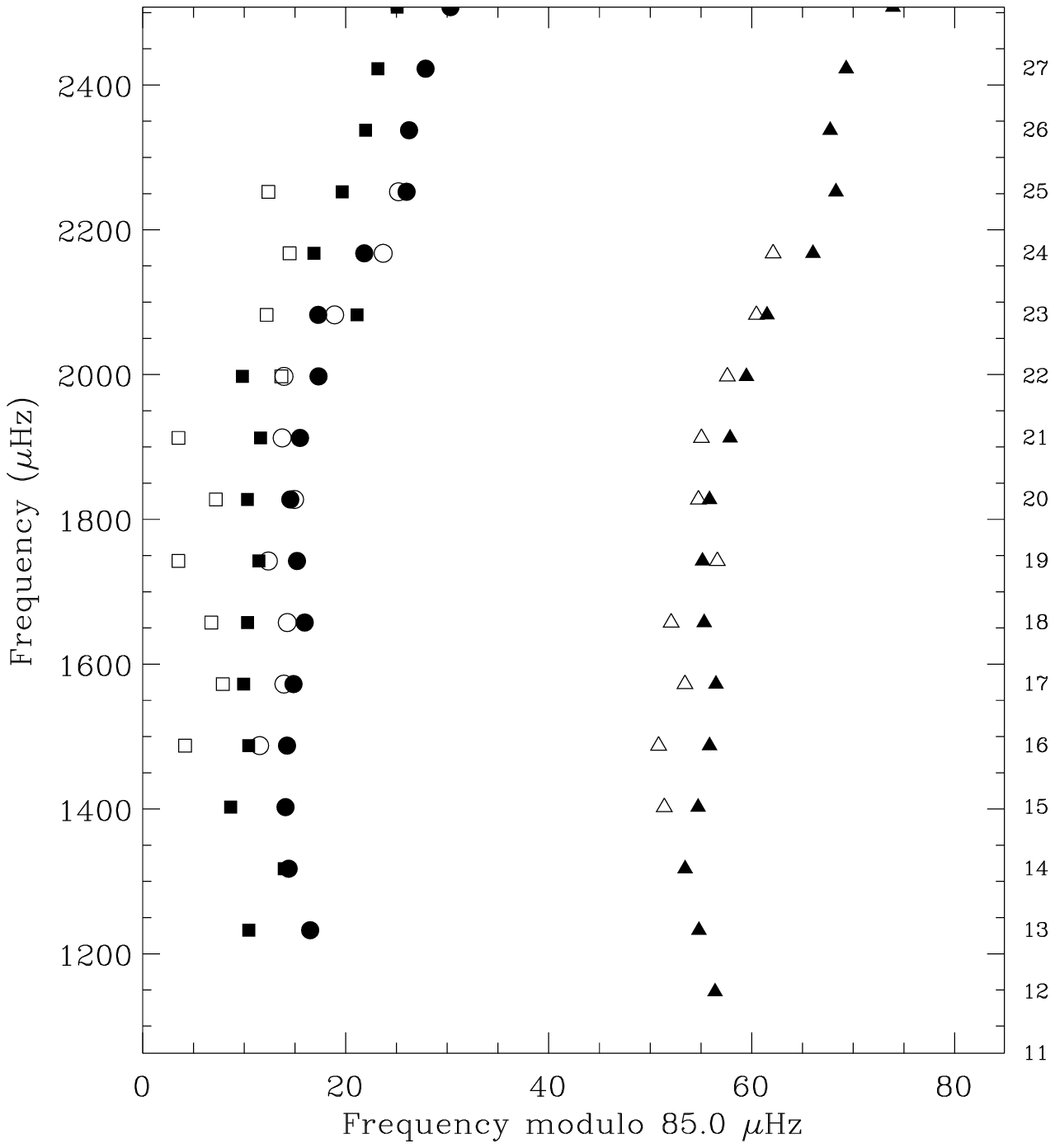}\\
\bigskip
\includegraphics*[width=72mm]{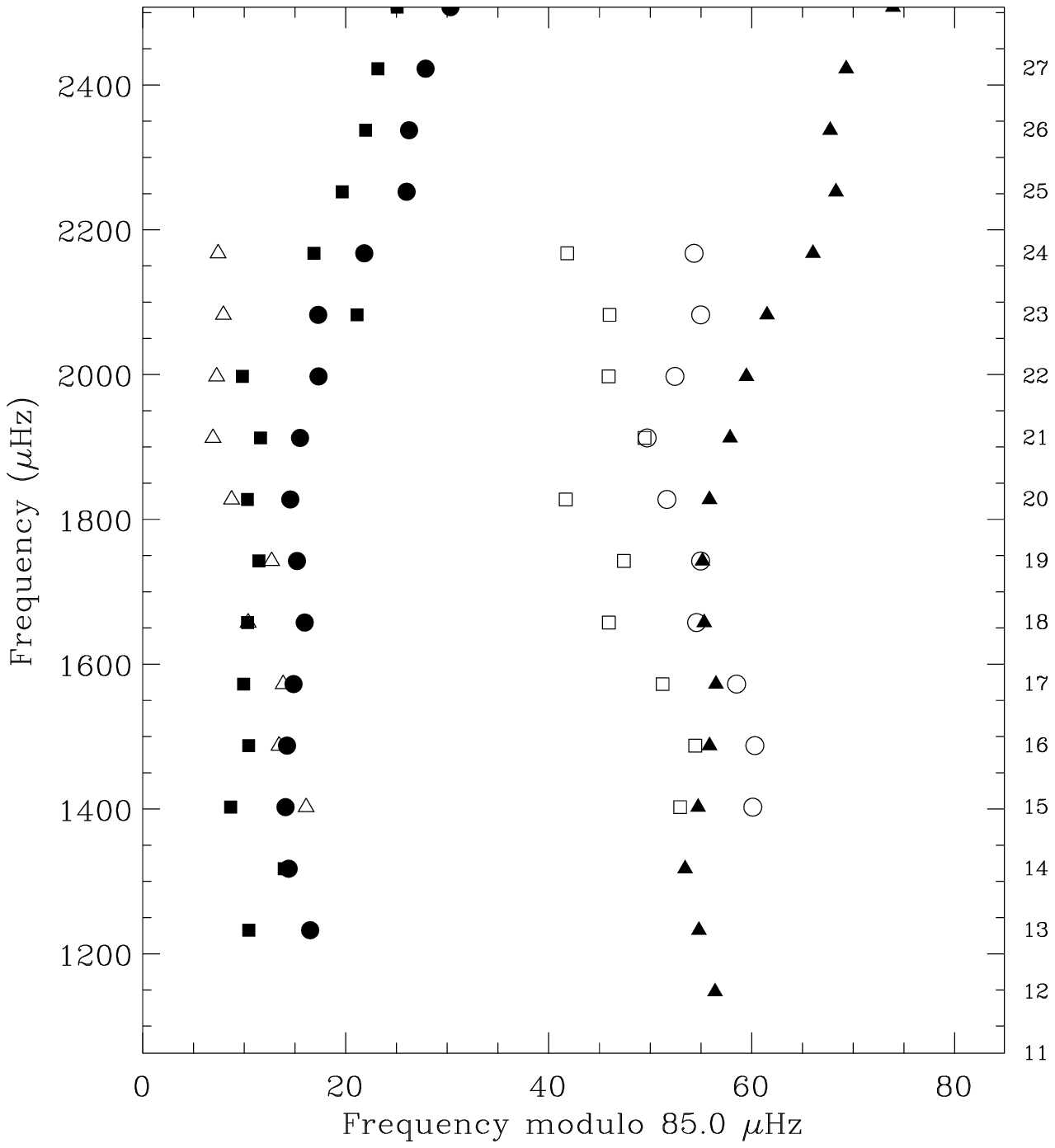}
\hfill
\includegraphics*[width=72mm]{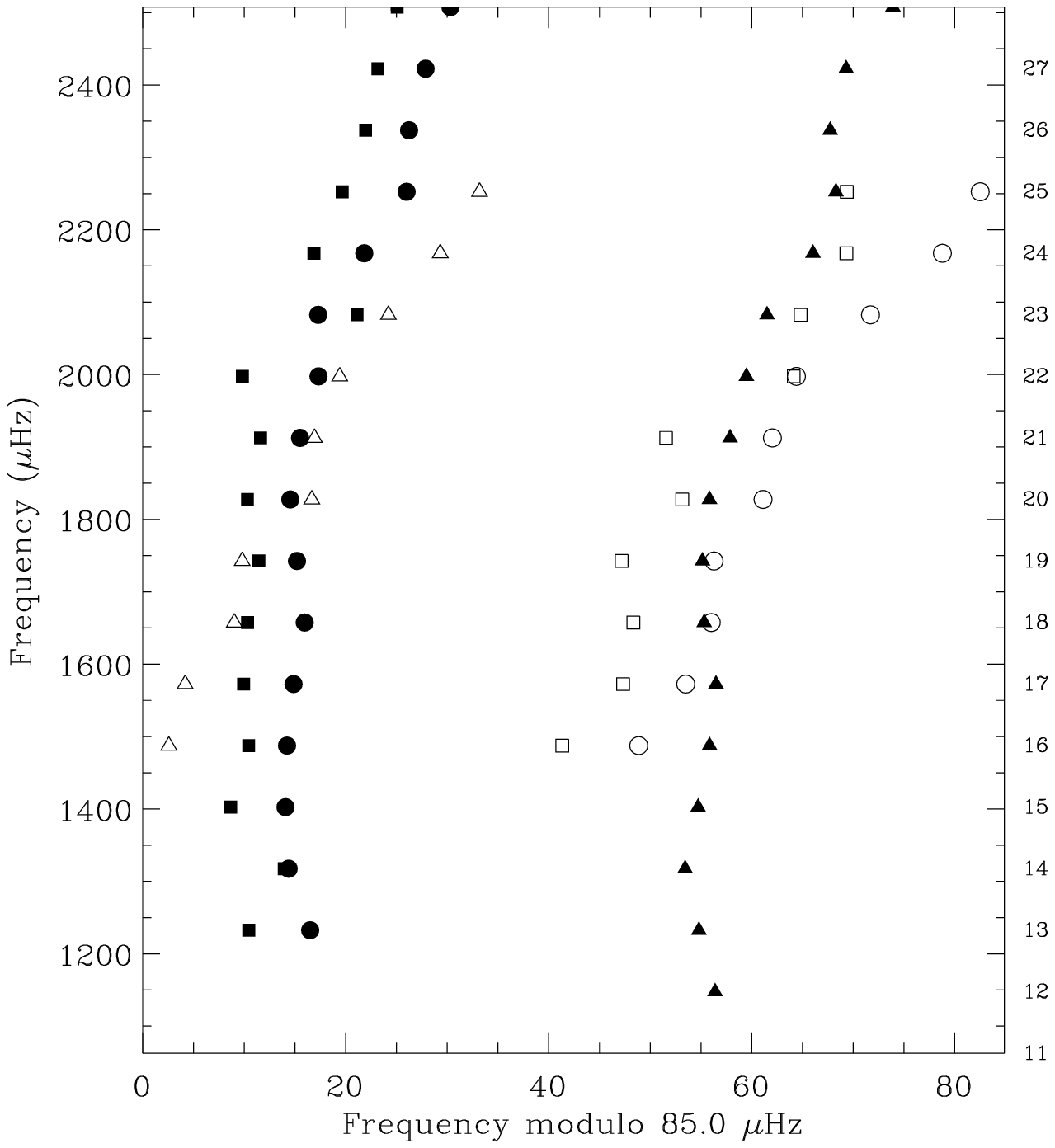}
\caption{Scaled \'echelle diagrams showing two CoRoT F-type targets that
 have ambiguous identifications.  The filled symbols are oscillation
 frequencies for HD~49933 from the revised identification
 \citep[Scenario~B]{BBC2009}. The open symbols are frequencies for
 HD~181420 \citep[Scenario~1]{BDB2009} after multiplying by 1.144 (upper
 panel), 1.115 (lower left) and 1.173 (lower right).  Symbol shapes
 indicate mode degree: $l=0$ (circles), $l=1$ (triangles), and $l=2$
 (squares).}
\label{fig:corot1}
\end{figure}

Using scaled \'echelle diagrams, together with the quite reasonable
assumption that $\epsilon$ varies slowly with stellar parameters, we might
hope to be able to tie these stars together.  Figure~\ref{fig:corot1} shows
how this works for two CoRoT targets, HD~49933 and HD~181420.  In all three
panels, the filled symbols show Scenario~B for HD~49933 \citep{BBC2009}.
The open symbols show Scenario~1 for HD~181420 \citep{BDB2009} with three
different scaling factors.  In the upper panel, the scaling factor was
chosen to align the ridges as closely as possible, \new{again using trial
and error,} and we indeed see a good match.  However, we should check
whether shifting one star by half an order also produces a match.  This
requires changing the scaling factor by $0.5/n_{\rm max}$, which is about
2.5\% in this case.  This is shown in the lower two panels of
Figure~\ref{fig:corot1}, where the scaling factor has been changed in both
directions by this amount and then fine-tuned to align the ridges.  Neither
of these match as well as the upper panel, giving us confidence that
Scenario~B for HD~49933 is equivalent to Scenario~1 for HD~181420.  That
is, the two scenarios are either both correct or both wrong.

\begin{figure}[t]
\centering
\includegraphics*[width=90mm]{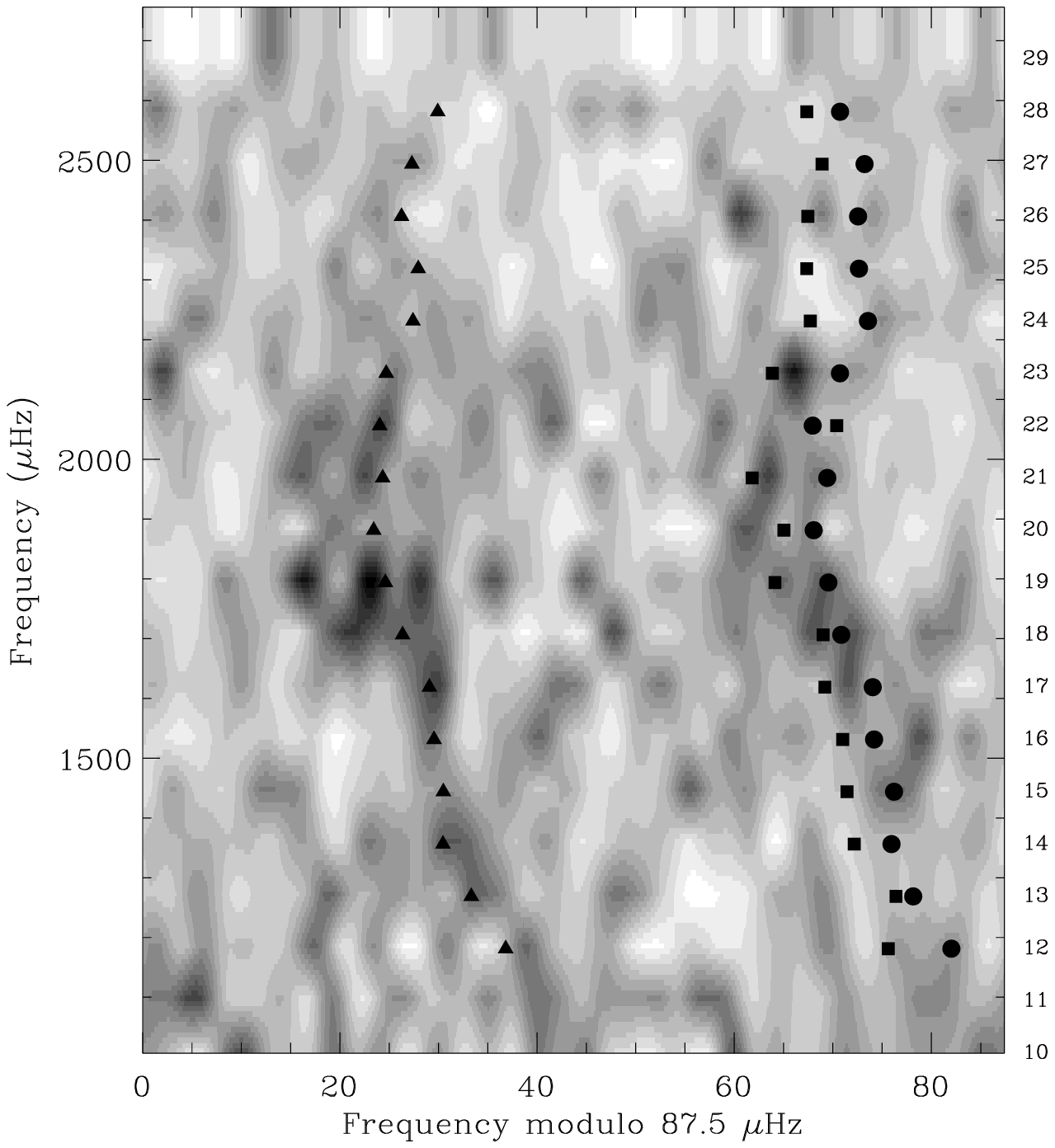}
\caption{Scaled \'echelle diagram comparing HD~49933 with another CoRoT
F-type target that has an ambiguous identification.  The greyscale is the
power spectrum of HD~181906 \citep{GRS2009}, smoothed to a FWHM of
3\,\muHz.  The filled symbols are oscillation frequencies for HD~49933 from
the revised identification \citep[Scenario~B]{BBC2009} after multiplying by
1.011.  Symbol shapes indicate mode degree: $l=0$ (circles), $l=1$
(triangles), and $l=2$ (squares).}
\label{fig:corot2}
\end{figure}

The third problematic CoRoT target mentioned above, which has a
significantly lower signal-to-noise ratio, is HD~181906 \citep{GRS2009}.
The power spectrum of this star is shown as the greyscale in
Figure~\ref{fig:corot2}.  Overlaid with filled symbols are the oscillation
frequencies for HD~49933 from the revised identification
\citep[Scenario~B]{BBC2009} after multiplying by 1.011.  There is good
agreement between the stars and, using HD~49933 as a guide, we are able to
follow the $l=1$ ridge of HD~181906 down to quite low frequencies.
Examining the two possible identifications proposed for HD~181906 by
\citet{GRS2009}, we can identify Scenario~B for that star with Scenario~B
for HD~49933.  Once again, either both are correct or both are wrong.

\begin{figure}[t]
\centering
\includegraphics*[width=90mm]{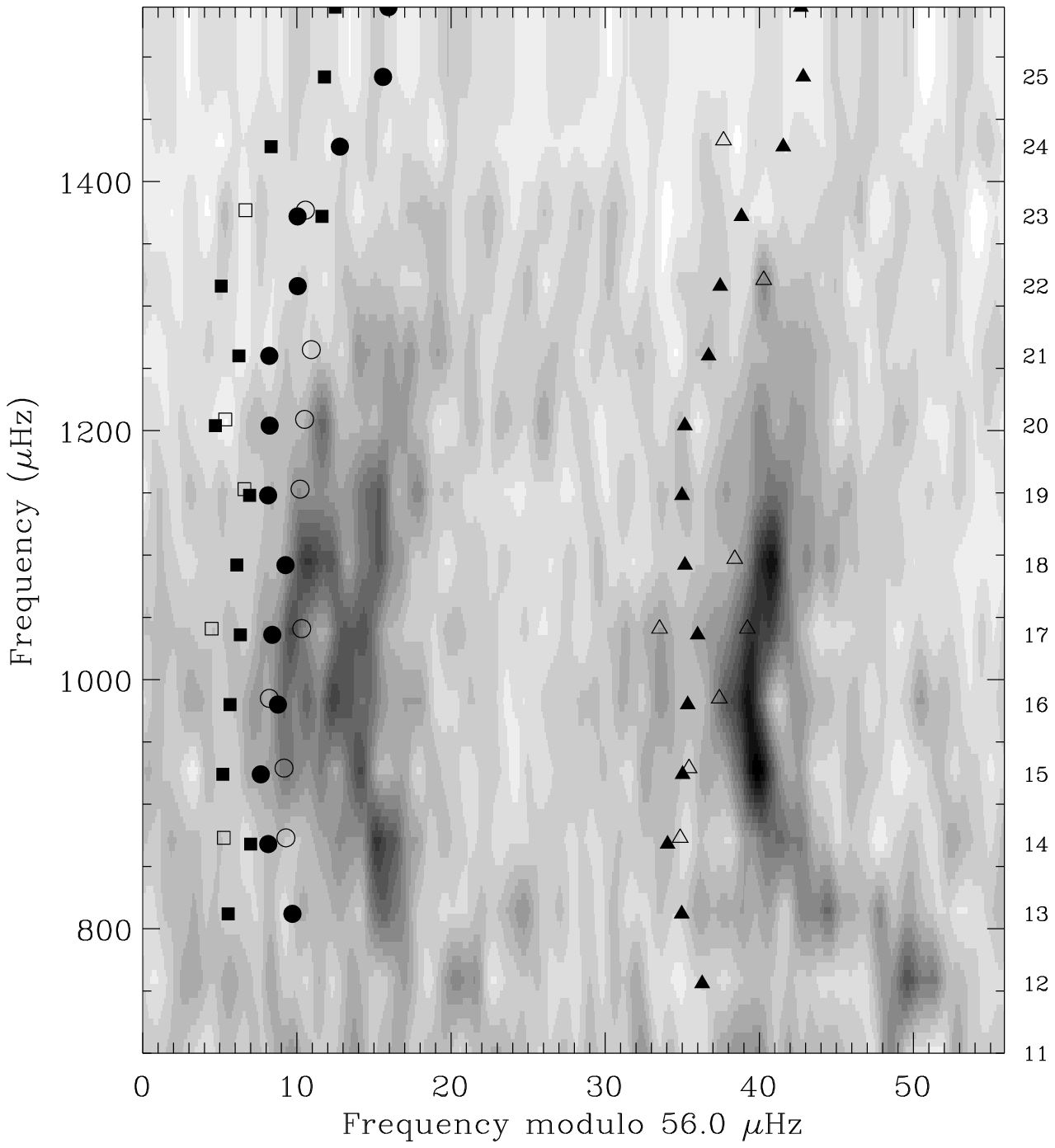}
\caption{Scaled \'echelle diagram to establish the correct identification
 for F-type stars.  The greyscale is the power spectrum of HD~49385
 \citep{DBM2010}, smoothed to a FWHM of 1\,\muHz.  The filled symbols are
 oscillation frequencies for HD~49933 from the revised identification
 \citep[Scenario~B]{BBC2009} after multiplying by 0.658.  Open symbols are
 frequencies for \eboo\ \citep{KBB2003} after multiplying by 1.390.  Symbol
 shapes indicate mode degree: $l=0$ (circles), $l=1$ (triangles), and $l=2$
 (squares).}
\label{fig:corot3}
\end{figure}

Having linked these three F-type CoRoT targets, all of which have quite
similar values of \Dnu, we would clearly like to confirm the
identifications by tying them to other stars whose identifications are
secure.  We do this in Figure~\ref{fig:corot3}.  The greyscale shows the
power spectrum of the CoRoT target HD~49385 ($\Dnu\approx 56\,\muHz$;
\citealt{DBM2010}), for which the $l=0$ and~2 ridges are clearly resolved.
The open symbols show frequencies for \eboo\ ($\Dnu\approx 40\,\muHz$)
after multiplying by 1.390, with mode identifications that were verified by
three sets of observations \citep{KBV95,KBB2003,CEB2005} Finally, the
filled symbols show once again the revised identification for HD~49933
\citep[Scenario~B]{BBC2009}, this time multiplied by 0.658.  These scaling
factors differ from unity by more than any others we have considered (since
\Dnu\ covers a bigger range).  Despite this, we see good alignment of the
ridges (Method~1) that gives a consistent picture.  Interestingly, HD~49385
shows significantly more curvature at the lowest orders than the other two
stars.

To summarise, we conclude that the correct identifications are: Scenario~B
for HD~49933 \citep{BBC2009}, Scenario~1 for HD~181420 \citep{BDB2009} and
Scenario~B for HD~181906 \citep{GRS2009}.  The first two of these agree
with the conclusions of \citet{M+A2009}, which were based on
autocorrelation analysis of the time series.

\section{Conclusions}

We have described a method for scaling oscillation frequencies and
displaying two or more stars on a single \'echelle diagram.  {Assuming that
two stars are sufficiently similar to be homologous, the diagram} can be
used to infer the ratio of their mean densities very precisely, without
reference to models.  In addition, data from the star with the better
signal-to-noise ratio can be used to confirm weaker modes and reject
sidelobes in data from a second star.  A very important application is to
provide a solution to the problem of ridge identification in F-type stars
observed by CoRoT, as discussed in Section~\ref{sec:f-stars}, {and we have
successfully applied the method to Procyon \citep{BKC2010}}.  Another
application is to apply what might be called ensemble asteroseismology to
the very large samples of stars being observed by the CoRoT and Kepler
space missions.  The results of applying this technique to red giants
observed with Kepler are described by \citet{BHS2010}.

\acknowledgements{ This work was supported financially by the Australian
Research Council and the Danish Natural Science Research Council.  We are
very grateful to Rafael Garc\'\i{}a for providing power spectra from CoRoT
observations in electronic form, and to S\'ebastien Deheuvels, Eric Michel
and colleagues for allowing us to show CoRoT results for HD~49385 in
advance of publication.  We thank Bill Chaplin and Dennis Stello for
encouragement and useful discussions.}

\end{document}